\documentstyle[preprint,aps,prb]{revtex}

\begin{document}
\title{Spin-Polarization Response Functions in High-Energy \\ 
($\overrightarrow{e}, e^{\prime}\overrightarrow{p}$) Reactions}
\author{Hiroshi Ito
\thanks{Also Department of Physics, the George Washington University, 
Washington, DC 20052}, 
S. E. Koonin, and Ryoichi Seki
\thanks{
Also Department of Physics and Astronomy, California State University,
Northridge, Northridge, CA 91330}
}
\address{W. K. Kellogg Radiation Laboratory, California Institute 
of Technology\\ Pasadena, CA 91125 \\
}
\date{\today}
\maketitle

\begin{abstract}
Spin-polarization response functions are examined for high-energy 
$(\overrightarrow{e},e^{\prime }\overrightarrow{p})$ reaction by computing 
the full 18 response functions for the proton kinetic energy $T_{p'}=$ 0.515 GeV 
and 3.179 GeV with an $^{16}O$ target. The Dirac eikonal formalism 
is applied to account for the final-state interactions.  The formalism is 
found to yield the response functions in good agreement with those calculated 
by the partial-wave expansion method at 0.515 GeV.  
We identify the response functions that depend on the spin-orbital potential 
in the final-state interactions, but not on the central potential. 
Dependence on the Dirac- or Pauli-type current of the nucleon is investigated 
in the helicity-dependent response functions, and the normal-component 
polarization of the knocked-out proton, $P_n$, is computed.
\end{abstract}

\newpage

\section{Introduction}

In a hard $(e,e'p)$ reaction, involving a few GeV/c 
or larger momentum transfer, the knocked-out proton experiences a strong, 
final-state interaction because of the large $pN$ cross sections (30-45 mb) 
corresponding to a short mean-free path of about 1.5 fm. 
Perturbative quantum chromodynamics suggests, however, the possibility 
of color transparency\cite{CTTH1,CTTH2}, in which the knocked-out proton 
undergoes little final-state interaction in the hard $(e,e^{\prime }p)$.  
The knocked-out proton would be
small (about the inverse of the momentum transfer) and 
color singlet, and would interact weakly with the other nucleons in
the nucleus through the color Van der Waals mechanism.  
This possibility has received much attention 
theoretically\cite{CTTH3,CTTH4,CTTH5,CTTH6,CTTH7} and 
experimentally\cite{CTEX1,CTEX2,CTEX3,CTEX4}.

Response functions from the $(e,e' p)$ reaction are affected 
greatly by the final-state interaction of the knock-out proton.  
Once the initial nuclear 
wave function is known (or assumed to be known), the response functions 
provide information of the final-state interaction, or the propagation 
of the knocked-out proton in nuclei.  
Polarization measurements 
in the $(\overrightarrow{e},e'p)$ and 
$(\overrightarrow{e},e' \overrightarrow{p})$ 
can provide detailed information on the process 
through the polarization response functions.  The polarization 
measurements in the GeV region are thus of great interest, 
and are planned to be carried out 
at the Thomas Jefferson National Accelerator Facility 
(TJNAF)\cite{GLAS,SAHA}. 

Theoretical investigation of the polarization response functions has been 
focused on the low proton energy of several hundred MeV or 
less~\cite{WALLY1,WALLY2,WALLY3}.  In the GeV 
region, though a few calculations have been carried out for the response 
functions of $(\overrightarrow{e},e^{\prime }p)$ 
in the last few years~\cite{GREEN1,GREEN2,ALDER}, 
no calculation is yet available for 
$(\overrightarrow{e},e^{\prime }\overrightarrow{p})$. 
 
In this paper we report the first calculation of the full set of the eighteen 
spin response functions for 
$(\overrightarrow{e},e^{\prime }\overrightarrow{p})$ 
in the GeV region, by incorporating spin-dependent, final-state interactions.  
We do not address the issue of the color transparency, but we   
calculate the response functions for the proton 
from different nuclear shell orbits and investigate 
their dependence on the spin-orbit interaction and the proton 
form factors.  We also discuss briefly the response functions of 
$(\overrightarrow{e},e^{\prime }\overrightarrow{n})$.

As in the works in low energies~\cite{WALLY1,WALLY2,WALLY3}, 
we employ the Dirac formulation for the bound-state 
wave functions and as in the previous works  
in the GeV region~\cite{GREEN1,GREEN2,ALDER}, we apply the Dirac eikonal 
formalism to the knocked-out proton wave function in the final state.
As in previous works, we neglect some physically important effects 
such as those that are due to off-shell effects and the current conservation.  
We do it in this exploratory work so as to establish bench-mark 
results, which could be compared with more refined calculations in future.

Though these effects are expected to be by no means negligible in the GeV 
region, the physics tends to become considerably simpler in comparison to that 
in the low-energy region:  The Dirac eikonal formalism has been successfully 
applied to the spin-asymmetry analysis of the proton-nucleus elastic 
scattering for the proton energy of 0.515 GeV\cite{AMADO}.
In the last few years, the same formalism has been applied to 
the $(\overrightarrow{e},e^{\prime }p)$ reaction 
in the GeV region\cite{GREEN1,GREEN2,ALDER}.  Later in this work, 
we explicitly demonstrate that the eikonal formalism is valid in the 
calculation of the response functions by comparing  
with the partial-wave decomposition method at 0.515 GeV.  
Our demonstration disagrees with the result of Ref.\cite{BIAN}, 
but agrees with its more recent result\cite{BIAN2}.

Being consistent with the Dirac eikonal description of the knocked-out proton,
we use the Hartree mean-field wave function of the Walecka model\cite{SEROT}
for the bound-state proton.  We thus neglect the nuclear correlation 
throughout this work.  Though the significance of the correlation effects 
on the high-energy $(e,e^{\prime }p)$ reactions in debate\cite{BEN,SEKI}, 
the effects appear to be small, once the other effects such as 
the finite range of the proton-nucleon interactions is included\cite{SEKI}. 

In Section II we review briefly the formalism for the 
$(\overrightarrow{e},e^{\prime } \overrightarrow{p})$ 
reaction and the Dirac eikonal method.  In Section III,
the numerical results of the 18 spin-dependent response
functions are presented, together with the examination of the
role of the spin-orbit force and the dependence on the structure of the
electromagnetic current operator.  The summary and the conclusion are given 
in Section IV.

\section{ Quasi-elastic Electron Scattering Formalism}

\subsection{Spin-Dependent Response Functions}

In this work, we follow the convention and notations of the 
$(\overrightarrow{e},e^{\prime } \overrightarrow{p})$ kinematics, 
which were used by Picklesimer and Van Orden\cite{WALLY2}. 
For convenience, the various kinematical quantities are illustrated in Fig.~1. 
The definition of the quantities are as follows:
The four momenta of the incoming and the outgoing electron are denoted as $k$ 
and $k^{\prime }$, respectively; 
the photon momentum is $q=k-k^{\prime }$ with $q^2\equiv q_0^2-{\bf q}^2<0$ 
(space-like); and the four momentum of the knocked-out proton is $p^{\prime }$.
We also denote $e$, $m_e$, and $M$ to be the electron charge, 
the electron mass, and the nucleon mass, respectively, 
and $E_{{\bf p^{\prime }}}=({\bf p^{\prime }}^2+M^2)^{1/2}$ 
to be the on-shell energy of the proton. 
We follow the Bjorken-Drell convention\cite{BDREL} of gamma matrices and 
Dirac spinors, in which the normalization condition is  
$\overline{u}(k,s) u(k,s)=1$ for the Dirac plane waves. 

In the following, we sketch the formalism on which our calculation is based. 
The formalism is of the standard, as described in Ref. (\cite{WALLY2}), but 
since it is rather involved, 
we wish to present it here for the sake of specifying notations and 
of clarifying the approximations involved in the quantities we 
calculate.

We assume 1) that the interaction between a proton in the nucleus and 
the electron is the one-photon exchange, and 
2) that the nuclear current consists of one-body currents.
We can then write the $(\overrightarrow{e},e^{\prime }\overrightarrow{p})$ 
cross section for $h$ and $\widehat{{\bf s}}$, the initial electron helicity 
and the spin polarization of the knocked-out proton, respectively, as

\begin{equation}
\left( \frac{d^3\sigma }{dE_{{\bf k}^{\prime }}d\Omega _{{\bf k}^{\prime
}}d\Omega _{{\bf p}^{\prime }}}\right) _{h,\widehat{{\bf s}}}=\frac{M|{\bf p}%
^{\prime }|}{(2\pi )^3}\left( \frac{d\sigma }{d\Omega _{{\bf k}^{\prime }}}%
\right) _{Mott}\sum_a\int dE_{{\bf p}^{\prime }}|{\cal M}_a|^2\delta (E_{%
{\bf p}^{\prime }}-q^0-M+\varepsilon _a),  
\end{equation}
\noindent 
summing over the occupied nuclear shell-orbits ($a$'s) in the single-particle 
description of the nucleus. ($\varepsilon _a$ is the binding energy in  
the $a$ shell.)  Here, the Mott cross section is 

\begin{equation}
\left( \frac{d\sigma }{d\Omega _{{\bf k}^{\prime }}}\right) _{Mott}
=\left( \frac{e^2\cos \frac \theta 2}
{8\pi |{\bf k}|\sin ^2\frac \theta 2}\right) ^2 , 
\end{equation}

\noindent 
where $\theta $ is the electron scattering angle.
The square of the transition amplitude for the knock-out proton
in the $a$-shell, $|{\cal M}_a|^2$, is written as a product of the
leptonic and nuclear tensors:

\begin{equation}
|{\cal M}_a|^2=\eta _{\mu \nu }W_a^{\mu \nu }. 
\end{equation}

\noindent 
The leptonic tensor is defined by

\begin{eqnarray}
\eta _{\mu \nu }
&=&m^2\sum_{s_e^{^{\prime }}}[\overline{u}(k,s_e)\gamma _\mu
u(k^{\prime },s_e^{\prime })][\overline{u}(k^{\prime },s_e^{\prime })\gamma
_\nu u(k,s_e)] \nonumber \\
&=&\frac 12(k_\mu k_\nu ^{\prime }+k_\nu k_\mu ^{\prime}-g_{\mu \nu }k
\cdot k^{\prime }
-ih\epsilon ^{\mu \nu \lambda \rho }k_\lambda ^{\prime }k_\rho ), 
\end{eqnarray}

\noindent 
where $s_e$ and $s'_e$ are the initial and final spins of the  
electron, respectively, and $\epsilon ^{\mu \nu \lambda \rho }$ is 
an antisymmetric, fourth-rank tensor.  Note that  
the electron mass is neglected in the second step of Eq. (4).

The nuclear tensor $W_a ^{\mu \nu }\equiv $ 
$W_a^{\mu \nu }(q ; {\bf p'}, \widehat{{\bf s}})$ depends on $q$, ${\bf p'}$,  
and $\widehat{{\bf s}}$, as well as on the quantum number of the $a$-shell 
orbit, and is written in terms of the matrix element of the nuclear current 
operator $J^\mu$,

\begin{equation}
W_a^{\mu \nu }(q ; {\bf p'}, \widehat{{\bf s}})
=\sum_{j_z} J_{a',\widehat{{\bf s}}}^{\mu \dagger}(q,{\bf p'})
             J_{a',\widehat{{\bf s}}}^\nu (q,{\bf p'}), 
\end{equation}
\noindent
where $a'$ is the quantum number of the proton (that is to be knocked out) 
in the $a$-shell, including $j_z$, the z-component of its total angular 
momentum.  The matrix element of $J_{\mu}$ is given by 

\begin{equation}
J_{a',\widehat{{\bf s}}}^\nu (q,{\bf p'})
=\langle\psi _{{\bf p}^{\prime },\widehat{{\bf s}}}^{(-)}
\Psi _F(A-1,a')|j^\nu (q)|\Psi _I(A)\rangle. 
\end{equation}

\noindent 
Here, $\psi _{{\bf p}^{\prime },\widehat{{\bf s}}}^{(-)}$ is the
scattered wave function of the knocked-out proton that satisfies the incoming
boundary condition.  $\Psi _I(A)$ is the initial, ground-state nuclear wave 
function, and $\Psi _F(A-1,a')$ is the final-state nuclear wave function 
with one hole that carries the quantum number $a'$.  $j^\nu (q)$ is the 
one-body current operator to be specified shortly. 

We introduce a M\"{o}ller-type operator, $\Omega^{(-)}$, that converts 
the Dirac plane wave to the distorted wave with the incoming boundary 
condition,  

\begin{equation}
\psi _{{\bf p}^{\prime },\widehat{{\bf s}}}^{(-)}
= \Omega^{(-)} u_{{\bf p}^{\prime },\widehat{{\bf s}}}^{(-)}.
\end{equation}
Note that $\Omega^{(-)}$ is not unitary,
as seen explicitly in Subsection II B.  
Equation (7) now allows us to write the nuclear tensor 
as the diagonal element of the the Dirac plane-wave spinor basis,  
$|u_{{\bf p}^{\prime },\widehat{{\bf s}}}^{(-)}>$:  

\begin{equation}
W_a^{\mu \nu }(q \; {\bf p'},\widehat{{\bf s}})
= Tr [ P_{\widehat{{\bf s}}} ({\bf p'}) \cdot 
\omega_a^{\mu \nu } (q)].
\end{equation}
Here, the spin-projection operator $P_{\widehat{{\bf s}}} ({\bf p'})$ 
is defined in terms of the Dirac plane-wave spinors as

\begin{eqnarray}
P_{\widehat{{\bf s}}} ({\bf p'})
&=& |u_{{\bf p}^{\prime },\widehat{{\bf s}}}^{(-)}\rangle
     \langle\overline{u}_{{\bf p}^{\prime },\widehat{{\bf s}}}^{(-)}| \\
&=&(\frac{\not{p'}+M}{4M})(1+\gamma^5\!\not{s}),
\end{eqnarray}
where the space-like, spin four vector $s^\mu$ is orthogonal 
to the momentum four vector of the knocked-out proton and 
is normalized to unity.  $s^\mu$ is related to the spin vector 
in the rest frame of the proton, $\widehat{{\bf s}}$, as 

\begin{equation}
s = \left( \frac{\widehat{{\bf s}} \cdot {\bf p'}}{M} , \widehat{{\bf s}}
+ \frac{\widehat{{\bf s}} \cdot {\bf p'}}{ M ( E_{\bf p'} + M )} 
{\bf p'}\right).
\end{equation}
$\omega_a ^{\mu \nu } (q)$ is the nuclear tensor in the Dirac 
plane-wave spinor space,  
\begin{eqnarray}
\omega_a ^{\mu \nu } (q) 
&=& \sum_{j_z} \widetilde{\Omega^{(-)}} 
\langle\Psi _F(A-1,a')|j^\nu (q)|\Psi_I(A)\rangle
\langle\Psi_I(A)|j^{\mu \dag} (q)|\Psi_F(A-1,a')\rangle \Omega^{(-)} \\
&=& s(a) \widetilde{\Omega^{(-)}} j^\nu (q) \sum_{j_z} |\psi _{a'}\rangle
\langle\psi _{a'}|j^{\mu \dag} (q) \Omega^{(-)} .
\end{eqnarray}
\noindent 
Here, $\psi_{a'}$ is the single-particle wave function of the proton 
in the $a$-th shell, $s(a)$ is its spectroscopic factor, and 
$\widetilde{\Omega^{(-)}}$ is the transpose of $\Omega^{(-)}$.  

As we define $\widehat{{\bf s}}$ in the rest frame of the proton, 
we decompose the trace in Eq.~(8) in terms of 
the spin-polarization response functions using the (right-handed) 
coordinate system in that frame.  We write the basis vectors 
of the coordinate system as 
($\widehat{{\bf n}} , \widehat{{\bf l}}, \widehat{{\bf t}}$).  
The spin-polarization is projected onto these vectors as 
${\cal S}_n=\widehat{{\bf n}}{\bf \cdot }\widehat{{\bf s}}$, 
${\cal S}_l=\widehat{{\bf l}}{\bf \cdot }\widehat{{\bf s}}$, 
and ${\cal S}_t=\widehat{{\bf t}}{\bf \cdot }\widehat{{\bf s}}$.
When the trace in Eq.~(8) is expressed in terms of these spin projections, 
the spin-polarization response functions, various $R^n$, $R^l$, and $R^t$, 
emerge in the coefficients of the spin projections, as seen below.

The differential cross section of the 
$(\overrightarrow{e},e^{\prime }\overrightarrow{p})$ 
reaction ejecting a proton with $h$ and $\widehat{{\bf s}}$ 
is now written in its full form,

\begin{eqnarray}
\left( \frac{d\sigma }{dE_{{\bf k}^{\prime }}d\Omega _{{\bf k}^{\prime }}
d\Omega _{{\bf p}^{\prime }}}\right) _{h,\widehat{{\bf s}}} 
&=&\frac{1}{2}
\left( \frac{d\sigma }{dE_{{\bf k}^{\prime }}d\Omega _{{\bf k}^{\prime }}
d\Omega _{{\bf p}^{\prime }}}\right) _{h} 
+\left[\left( \frac{d\sigma }{dE_{{\bf k}^{\prime }}d\Omega _{{\bf k}^{\prime }}
d\Omega _{{\bf p}^{\prime }}}\right) _{h,\widehat{{\bf s}}} 
- \frac{1}{2}
\left( \frac{d\sigma }{dE_{{\bf k}^{\prime }}d\Omega _{{\bf k}^{\prime }}
d\Omega _{{\bf p}^{\prime }}}\right) _{h} \right]  \nonumber \\
&\equiv&\frac{1}{2}\sigma( h, 0 ) + \sigma( h,\widehat{{\bf s}} ), 
\end{eqnarray}
where $\sigma( h, 0 )$ is the differential cross section 
for $(\overrightarrow{e},e'p)$ and is given by

\begin{eqnarray}
\sigma( h, 0 )
&=&\frac{M|{\bf p}^{\prime }|}{(2\pi )^3}\left( \frac{d\sigma }
{d\Omega _{{\bf k}^{\prime }}}\right)_{Mott}  \nonumber \\
&\cdot& \{v_L R_L+v_T R_T+v_{TT} R_{TT}\cos 2\beta 
+v_{LT} R_{LT}\sin \beta +hv_{LT^{\prime }} R_{LT^{\prime }}\cos \beta \}. 
\end{eqnarray}
$\sigma( h,\widehat{{\bf s}})$ is the polarized part of 
the ($\overrightarrow{e}, e^{\prime}\overrightarrow{p}$) differential 
cross section and is given by 

\begin{eqnarray}
\sigma( h,\widehat{{\bf s}})
&=&\frac{M|{\bf p}^{\prime }|}{2(2\pi )^3}
\left( \frac{d\sigma }
{d\Omega _{{\bf k}^{\prime }}}\right)_{Mott} \nonumber \\
&\cdot&\{[v_L R_L^n +v_T R_T^n +v_{TT} R_{TT}^n\cos 2\beta
   +v_{LT} R_{LT}^n\sin \beta + h v_{LT^{\prime }} R_{LT'}^n \cos \beta]
{\cal S}_n \nonumber \\ 
& &+[v_{TT}R_{TT}^l\sin 2\beta +v_{LT} R_{LT}^n\cos \beta  
   + h (v_{LT^{\prime }} R_{LT'}^n \sin \beta+v_{TT^{\prime }} R_{TT'}^n)]
{\cal S}_l \nonumber \\ 
& &+[v_{TT}R_{TT}^t\sin 2\beta +v_{LT} R_{LT}^t\cos \beta  
   + h (v_{LT^{\prime }} R_{LT'}^t \sin \beta+v_{TT^{\prime }} R_{TT'}^t)]
{\cal S}_t \} \nonumber \\ 
&\equiv& N_n{\cal S}_n+N_l{\cal S}_l+N_t{\cal S}_t,
\end{eqnarray}

\noindent 
where $\beta$ is the azimuthal angle of ${\bf p'}$ 
as illustrated in Fig.~1; and $v$'s 
($v_L, v_T, v_{TT}, v_{LT}, v_{LT^{\prime }},$ and $v_{TT^{\prime }}$)
are kinematic factors, depending only on $\theta, {{\bf q}^2},$ and $q^2$.
For completeness, in the Appendix we list the relations between the response 
functions 
and the nuclear tensor, and the explicit forms of the kinematic factors. 

In the experiments planned at the TJNAF, simplified kinematics is applied 
to reduce the number of the response functions involved: 
The in-plane kinematics of $\beta=n\pi$ is used for polarized 
beams\cite{GLAS} and for unpolarized ($h = 0$) beams\cite{SAHA}.
In the latter case, the induced polarization yields the helicity-independent 
(nonzero) normal polarization component.  The differential cross section 
for this $(e,e^{\prime } \overrightarrow{p})$ is written in terms 
of the preceding $\sigma(h, 0)$ and $N_n$ (but setting $\beta=n\pi$) as

\begin{equation}
\left( \frac{d\sigma }{dE_{{\bf k}^{\prime }}d\Omega _{{\bf k}^{\prime }}
d\Omega _{{\bf p}^{\prime }}}\right) _{h,\widehat{{\bf s}}} 
=\frac{1}{2}{\sigma( h, 0 )}_{\beta=n\pi}[1 + P_n],
\end{equation}
where 

\begin{equation}
P_n = {\left[ N_n / \sigma( h,0 ) \right]}_ {\beta = n\pi}.
\end{equation}
In Section II D, we discuss our numerical results of $P_n$.

In this work, we use the one-body current operator in free space,

\begin{equation}
{j}^\mu (q)=\gamma ^0\left[ F_1(q^2)\gamma ^\mu +i{\frac \kappa {2M}}%
F_2(q^2)\sigma ^{\mu \nu }q_\nu \right], 
\end{equation}
by neglecting off-shell effects involved in the current\cite{FORE}.
Different prescriptions for the off-shell extension of the current, as well
as for recovering the current conservation, are recently 
discussed\cite{POLLO} and will be commented on in Section IV.
In this work, we use the standard dipole function for the Dirac and 
the Pauli form factors $F_1(q^2)$ and $F_2(q^2)$(with $\kappa =1.79$), 
except when noted otherwise.

\subsection{Dirac Eikonal Approximation}

The initial- and final-state proton wave functions, $\psi _{a'}(\hbox{\bf r})$ 
and $\psi _{{\bf p'},s}^{(-)}(\hbox{\bf r})$ satisfy the
Dirac equation with the scalar potential $V_s$, and the vector potential 
$V_v. $ $\psi _{a'} (\hbox{\bf r})$ is the quantum-hadrodynamical 
wave function in the Hartree approximation\cite{SEROT}, and is expressed 
in the standard form\cite{BDREL},

\begin{equation}
\psi _{a'} (\hbox{\bf r})=\frac 1r\left( 
\begin{array}{c}
iG_{n,\kappa }(r)\Phi _{\kappa, j_z}(\Omega ) \\ 
-F_{n,\kappa }(r)\Phi _{-\kappa, j_z}(\Omega ) 
\end{array}
\right) 
\end{equation}

\noindent 
for the nuclear shell state $a$ with $a' = (n, j, l, j_z)$, where $j$ and $l$ 
are specified through a quantum number $\kappa $. 
The wave function is normalized to unity, and $\Phi _{\!\pm \kappa ,j_z}$ 
are the spin spherical harmonics for the solid angle, $\Omega$.

The continuum-state wave function of the proton with the momentum 
${\bf p}^{\prime}$ and the spin $s$ is expressed as

\begin{equation}
\psi_{{\bf p}^{\prime},s}=\left( 
\begin{array}{c}
u_{{\bf p}^{\prime},s} \\ 
w_{{\bf p}^{\prime},s} 
\end{array}
\right) ,
\end{equation}

\noindent 
where each component satisfies

\begin{eqnarray}
&&\left[ \frac{-\nabla ^2}{2M}+V_C
+V_{SO}({\bf \sigma }\cdot {\bf L}-i{\bf r}\cdot{\bf p}^{\prime}) \right] 
u_{{\bf p}^{\prime},s}
=\frac{{\bf p}^{\prime 2}}{2M} u_{{\bf p}^{\prime},s} \nonumber \\ 
&&w_{{\bf p}^{\prime},s}
=-\frac i{D(r)}({\bf \sigma }\cdot {\bf \nabla )}
u_{{\bf p}^{\prime},s}, 
\end{eqnarray}

\noindent 
where $D(r)=E+M+V_s(r)-V_v(r)$. Here, $V_C$ and $V_{SO}$ are the
central and spin-orbit potentials related to $V_s$ and $V_v$ by

\begin{eqnarray}
V_C(r)  &=& V_s+\frac EMV_v+\frac{V_s^2-V_v^2}{2M} \nonumber \\ 
V_{SO}(r) &=& \frac 1{2MD(r)}\frac 1r\frac d{dr}[V_v-V_s]. 
\end{eqnarray}

\noindent The solution of Eq.~(22) with the incoming boundary condition 
is given, in the eikonal approximation, by 

\begin{equation}
\psi _{{\bf p}^{\prime },s}^{(-)}({\bf r})=
\left(\frac{E_{{\bf p}^{\prime }}+M}{2E_{{\bf p}^{\prime }}}\right)^{1/2}
\left( 
\begin{array}{c}
1 \\ 
-iD(r)^{-1}({\bf \sigma \cdot \nabla }) 
\end{array}
\right) e^{i{\bf p}^{\prime }.{\bf r}}e^{iS({\bf r})}\chi _s. 
\end{equation}

\noindent Here, $S({\bf r})$ is the eikonal phase, 

\begin{equation}
S({\bf r})=\frac M{p^{\prime }}\int\limits_z^\infty dz^{\prime
}\{V_C(z^{\prime },{\bf b})+V_{SO}(z^{\prime },{\bf b})[{\bf \sigma \cdot
b\times p}^{\prime }{\bf -}ip^{\prime }z^{\prime }]\}, 
\end{equation}

\noindent where ${\bf r=}$ $z$ ${\bf e}_z+b{\bf e}_{\bot }$ with ${\bf e}_z$
and ${\bf e}_{\bot }$ being the longitudinal and transverse unit vectors along
the direction of ${\bf p}^{\prime}$.  In this work, we are 
interested in each contribution of the central and spin-orbit forces to the
18 spin-dependent response functions.  We implement
this by switching on and off $V_C$ and $V_{SO}$ in Eq. (22).

\section{Numerical Results}

We now describe the numerical results of the spin-dependent response functions
for the ($\overrightarrow{e},e^{\prime }\overrightarrow{p}$) reaction, taking 
$^{16}O$ as an example.  After establishing the accuracy of the eikonal 
approximation (in Subsection A), we illustrate the response functions 
and examine effects of the spin-orbital force (in Subsection B) and 
of the nucleon electromagnetic form factors (in Subsection C). 
We also show the normal-component polarization relevant to an 
experiment planned at TJNAF\cite{SAHA} (in Subsection D).  We present 
the results 
at two kinetic energies of the knock-out proton, $T_{p'}=0.515$ GeV and 
$3.179$ GeV.   These energies correspond to the extreme energies in 
the experiment\cite{SAHA}.  At the lower energy of $0.515$ GeV, 
we compare the response functions calculated by the eikonal and partial-wave 
decomposition methods.  Since no detailed 
phenomenological optical potential is available at these energies, we use the
optical potential in the lowest-order impulse approximation, the so-called 
$f\rho $-form, where the nuclear density $\rho $ is taken from the
Hartree mean-field nuclear wave function.  The description of this
method in Dirac formalism is elaborated in Ref.\cite{WALLACE} and is 
summarized in \cite{GREEN2}. 
We use the $pN$-scattering amplitudes from the phase-shift analyses of 
Ref.\cite{HOSHIZ} and Ref.\cite{HIGUCH} for $T_{p'}=0.515$ and $3.179$ GeV, 
respectively.

\subsection{Dirac Eikonal Approximation vs. Partial-Wave Decomposition Method
}

To compare the Dirac eikonal and partial-wave decomposition methods, 
we select ten representative response functions out of the full 18 functions, 
and show the results at $T_{p'}=0.515$ GeV $(|{\bf p}| = 1.113 GeV/c$) with 
$Q^2 \equiv -q^2 = 1 (GeV/c)^2$ in Fig. 2.   
The response functions are shown in the commonly used kinematics 
in the low energies, as a function of the magnitude of the recoil 
momentum of the residual nucleus, $|{\bf p}^{\prime} -{\bf q}|$, at a  
constant momentum transfer $|{\bf q}|$ (here, $|{\bf q}| = 1.113 GeV/c$).

The Dirac partial-wave decomposition method is fully described in 
Ref.\cite{WALLY1}, and the response functions by the method shown in Fig. 2 
are provided to us by J. W. Van Orden\cite{WPRI}.   The response functions 
in Fig. 2 by the two methods are obtained in the same kinematics,  
using the same input parameters together 
with the H\"ohler nucleon electromagnetic form factor\cite{HOHLER}. 
Figure 2 shows that the results by the two methods are quite close, within 
10\% at the peak for all response functions shown.  The exception is 
with $R_{TT}^n$, for which the discrepancy at the peak is larger (about 20\%). 
Note that a similar, relatively large ($\sim$ 20 \%) discrepancy is seen with 
one of the $t$-component response function, $R^{t}_{TT}$ (not shown here).

In order to solidify this comparison, we repeat the comparison 
at $T_{p'}=135 MeV$ 
and find the discrepancy to be much larger, 
typically of 30-40\%, and even larger (80 -- 100\%) for the transverse
   responses ($R^{t}_{TT}, R^{n}_{TT}$ and $R^{l}_{TT}$).
(We do not exhibit the $135 MeV$ results in order to keep the number 
of figures reasonable.)  As we go up to the GeV region, the number of 
the partial waves naturally increases, and the partial-wave decomposition 
method becomes more elaborate and eventually become impractical.
On the other hand, the eikonal becomes more accurate as 
the ratio of $T_{p'}$ and (the magnitude of) the $pN$ potential increases.  
Though we have no partial-wave decomposition result available to compare 
at the GeV region, we expect the eikonal method to be reasonably accurate. 
The Dirac eikonal method should be the practical, reasonably reliable method for examining the final-state
interaction in the high-energy $(\overrightarrow{e},e^{\prime }%
\overrightarrow{p})$ reaction.

\subsection{Spin-Orbit Force}

Figures 3 and 4 show the complete set of 18 spin-dependent
response functions for the proton knock-out from the $p_{1/2}$-shell with
the kinetic energy of $T_{p'}=0.515$ GeV (the same kinematics as that used 
in Subsection II B, $|{\bf p}^{\prime }|= |{\bf q}| = 1.133$ GeV/c). 
The response functions are calculated 
with and without the final-state interaction (that is, DWIA and PWIA, 
respectively.)  The DWIA responses are generally smaller in magnitude than 
the PWIA responses, as a consequence of the absorption in the final-state 
interaction.  The largest response function is $R_T$ among the unpolarized 
response functions, $R_{L,}R_T,R_{TT},R_{LT}$ and $R_{LT^{\prime}}$, and 
dominates the unpolarized cross section. 

The helicity-dependent response function, $R_{LT^{\prime }}$, vanishes 
in the absence of the final-state interaction and is a quantity useful 
for the investigation of the proton-flux attenuation by the final-state 
interaction.  At the parallel kinematics
(i.e., $|{\bf p}^{\prime }$-${\bf q}|=0$)$,$ $R_{TT},$ $R_{LT}$ and $%
R_{LT^{\prime }}$ vanish.  At $T_{p'}=0.135$ GeV, it was observed\cite{WALLY1} 
by the partial-wave decomposition calculation, that 
the sign of $R_{TT}$ changes by the inclusion of the final-state interaction 
for the proton knocked out from the $1p_{1/2}$-shell.  We find the same 
to occur at this energy and also at $T_{p'} = 3.197$ GeV.  
Figure 5 shows the response functions $R$'s and $R^n$'s for the proton 
knocked out from the $1p_{3/2}$-shell.  Here, the sign of $R_{TT}$ remains 
the same with the inclusion of the final-state interaction as is the case 
at $T_{p'}=0.135$ GeV\cite{WALLY1}.  
The response functions for the polarized proton in the ${\bf n,l}$ and 
${\bf t}$ directions are also shown in Fig.4, many of which vanish 
in the absence of the final-state interaction.  

Figures 6 and 7 illustrate the response functions for the proton knocked out 
from the $1p_{1/2}$-shell at $T_{p'}= 3.179$ GeV 
($|{\bf p}^{\prime }|=  4.024$ GeV/c)
with $|{\bf q}|= 4.024$ GeV/c, and $Q^2 = 6$ (GeV/c${)}^2$.  
The magnitude of the response functions 
at this energy is typically smaller by two orders of magnitude than those 
at $T_{p'}=0.515$ GeV.  This reduction is caused mostly by the 
$Q^2$ dependence of the nucleon electromagnetic form factor, 
the square of which contributes to the response functions.  
Clearly, a further increase in $Q^2$ that is expected in future 
experiments will reduce considerably the magnitude of the response functions. 

Note that in order to keep the number of figures reasonable, we have selected 
the figures to be presented in this work: We show the full set of the response 
functions for the proton knocked out 
from the $1p_{1/2}$-shell at $T_{p'}= 0.515$ and 
3.179 GeV, so that one could compare them with the lower-energy result at 
$T_{p'} = 0.135$ GeV in Ref.\cite{WALLY3}.  We also show the unpolarized and 
normal-component, polarized response functions ($R$'s and $R^n$'s, respectively)   
for the case of the $1p_{3/2}$-shell because of their greater 
contributions to $P_n$ and the spin-orbital effects 
than the $R^l$'s and $R^t$'s.

It is interesting to examine how the spin-dependent force in the final-state 
interaction affects the response functions.  
For this purpose, we repeated the calculation by omitting the spin-orbit force 
from the final-state interaction ($V_{SO}=0$). 
The resultant response functions are shown in dash lines  
in Figs.~3 - 9.  We see that the interesting sign change of $R_{TT}$
noted above can be attributed to the effect of the spin-orbit force, 
as clearly demonstrated in $R_{TT}$ of Fig.3.  As a consequence of the 
interference between the effects of the central and spin-orbital interactions,
the spin-orbital force {\it increases} the $TT$ component of the outgoing 
flux of the proton.  By comparing $R_{TT}$ of Fig. 3 and Fig. 6, we observe 
that this effect becomes relatively weaker as the energy increases.  

The response function for the normally polarized response state $R_T^n$ has 
a similar feature, but here, the plane-wave response and the response 
without the spin-orbital force vanish. 
That is, the central force does not affect $R_T^n$, but only 
the spin-orbital force does.
The situation is opposite in $R_{TT}^n$, in which the effect of 
the final-state interaction is dominated by the central force.  
The same features as described here are also seen in Fig. 5 in the case 
of the $1p_{3/2}$-shell. 

Finally, we note that the signs of response functions are generally opposite 
for the 1$p_{1/2}$- and $1p_{3/2}$-shell, except for the cases of 
$R_{L,}R_T,R_{LT}$, and $R_{LT^{\prime }}^n.$

\subsection{Electromagnetic Form Factors of The Nucleon}

We also examine the dependence of response functions on the
structure of the nucleon electromagnetic current.  Figure 8 illustrates 
the response functions with the Dirac-type current ($\gamma ^\mu$) only 
(obtained by setting 
$F_2(q^2)=0$ and $F_1(q^2)\neq 0$), in the case of the proton knock-out 
from the $1p_{1/2}$-shell at $T_{p'}=0.515$ GeV.  The response
functions with the Pauli current ($\sigma ^{\mu \nu }q_\nu $) only 
(obtained by setting $F_1(q^2)=0$ and 
$F_2(q^2)\neq 0)$ are shown in Fig.9.  Note that the response functions
shown in Fig. 3 correspond (roughly speaking) to the sum of these two 
($F_1$ and $F_2$), including the interference between them. 
We observe that these two types of the electromagnetic 
current are equally important for most of the response functions. 
Figures 8 and 9 also include similar calculations without the spin-orbital 
force in the final-state interaction.  We also observe the same feature 
in this case.

In the cases of the helicity-dependent response functions, 
$R_{LT^{\prime }}$, $R_{LT^{\prime}}^n$, and $R_{TT}$, 
the contributions of the Dirac-type and the Pauli-type currents have opposite 
signs, while the signs remain the same in the other response functions.  
The neutron has a net zero charge, and its Dirac
form factor is extremely small $(F_1\simeq 0)$, as is well-known from the fact
that the Sachs charge radius of the neutron is almost completely saturated 
by the magnetic radius.  The response functions shown in Fig. 9 are thus 
expected to be similar (sign-wise and magnitude-wise) to the response 
functions for 
the $(\overrightarrow{e},e^{\prime }\overrightarrow{n})$ reaction. 
We have confirmed this expectation by calculating the response functions 
for the $(\overrightarrow{e},e^{\prime }\overrightarrow{n})$ reaction 
with the realistic neutron form factors.  Note that we are neglecting  
the charge-exchange contribution to 
the $(\overrightarrow{e},e^{\prime }\overrightarrow{n})$ reaction, but 
the contribution is expected to be relatively small in the GeV 
energy region.  A further note on a more detailed feature: 
The helicity-dependent response functions, 
$R_{LT^{\prime }}$ and $R_{LT^{\prime }}^n$, have opposite signs in 
$(\overrightarrow{e},e^{\prime }\overrightarrow{n})$ and  
$(\overrightarrow{e},e^{\prime }\overrightarrow{p})$.

\subsection{Polarization of the Knocked-out Nucleon: $P_n$}

The normal-component polarization of the outgoing proton, $P_n$, can be
observed in the $(e,e^{\prime }\overrightarrow{p})$ reaction with an 
unpolarized electron beam\cite{SAHA}.   $P_n$ is expressed in terms of  
of the response functions as shown in Eqs. (15)-(18). 
Figure 10 illustrates $P_n$ for the proton knock-out from the 
the $1p_{1/2}$- and $1p_{3/2}$-shells at $T_{p'}=0.515$ GeV.  
In the absence of
the final-state interaction, the normal component of the spin-dependent
response functions $R_L^n,R_T^n,R_{TT}^n$ and $R_{LT}^n$ vanish, so that 
$P_n=0$ in the PWIA.  $P_n$ for the $1p_{1/2}$ shell is negative for 
$|{\bf p}^{^{\prime }} - {\bf q}| < 1.5$ fm${}^{-1}$,  
while $P_n$ for the $1p_{3/2}$-shell is positive  
for $|{\bf p}^{^{\prime }}- {\bf q}| < 1$ fm${}^{-1}$.  
The polarization induced only by the central force $V_C$ is
also shown in Fig. 10.  
Similar results for $T_{p'}=3.179$ GeV are shown in Fig.11.
The nuclear-recoil dependence of $P_n$ is similar at both energies, but 
its magnitude is considerably smaller (by more than 40\%) at $T_{p'}=3.179$ GeV
than at $T_{p'}=0.515$ GeV, even becoming comparable to the expected 
experimental accuracy $\Delta P_n\simeq 0.5$\cite{SAHA}.

The polarization of the outgoing proton $P_n$ is induced by the final-state
interaction, so it vanishes in the absence of the final-state interaction.
In fact, $P_n$ is insensitive to the structure of the electromagnetic current:  
Numerically we find $P_n$ for the two cases, $F_1(q^2)\neq 0$ with $F_2(q^2)=0$ 
and $F_2(q^2)\neq 0$ with $F_1(q^2)=0$, to be practically identical. 

We have also examined $P_n$ for the 
$(\overrightarrow{e},e^{\prime }\overrightarrow{n})$ and  
$(\overrightarrow{e},e^{\prime }\overrightarrow{p})$ reactions at 
different $T_{p'}$ from different shell-orbits.  $P_n$ for the two reactions 
are found to be almost identical, but as noted previously, our calculation 
does not include the charge-exchange interaction.  

\section{Discussion}

We comment on the two important effects that we have neglected in this work.

{\it The current conservation}.  
A DWIA calculation of the $(\overrightarrow{e},e^{\prime
}\overrightarrow{p})$ amplitude suffers from the violation of current
conservation.  The violation arises basically 
in the truncation of the many-body degrees 
of freedom by reduction to the one nucleon problem of the mean-field theory. 
It is also closely related to the treatment of the off-shell effects. 
  
The current conservation implies a constraint on the nuclear matrix elements of
the longitudinal and time components, $q^0J_{\alpha ,\widehat{{\bf s}}%
}^0(q)=|{\bf q}|J_{\alpha ,\widehat{{\bf s}}}^L(q).$  A quantity such as  
$(R_L-\widetilde{R_L}$) /$(R_L+\widetilde{R_L})$ would provide a measure of 
the violation\cite{WALLY1}.  Here, the longitudinal response function $R_L$ 
is calculated by the use of $J_{\alpha ,\widehat{{\bf s}}}^L(q)$, and 
the $\widetilde{R_L}$ is by the use of 
$q^0J_{\alpha ,\widehat{{\bf s}}}^0(q)/|{\bf q}|$. 
Though the quantity was found to reach nearly 40 \% 
at $T_{p'} = 135$ MeV\cite{WALLY1}, it has been estimated to be much less, 
$\leq 10\%$, for $T_{p'}> 0.515$ GeV\cite{GREEN2}.  
The latter high-energy estimate is 
comparable to other uncertainties in our calculation, such as those 
in the optical-potential parameters.  Note, however, the normal-component 
polarizations, which are important to $P_n$, for example, 
would be less affected by the 
nonconservation, because these quantities depend mostly on the transverse 
components that are not associated with the current conservation. 

{\it The off-shell effects}: 
The issue of the nonconserved current is complicated because of the 
off-shell effects because there is no unique way to recover the current 
conservation for the off-shell nucleon.  For example, other forms 
of the one-body current operator $j^{\mu}(q)$ that are equivalent to 
Eq. (18) by means of the Gordon decomposition are 
no longer equivalent\cite{FORE}.  Recently, in PWBA, 
the off-shell effects for $(e,e^{\prime} p)$ are estimated 
to be $\leq 10\%$ in the GeV region after the current conservation 
is imposed in various ways\cite{POLLO}.

From these, we suspect that the important physics neglected in this work 
could contribute appreciably.  Clearly, more refined work is needed 
to establish reliable results.

\section{Summary and Conclusion}

In this work, we have presented the first DWIA calculation of 
the spin-polarization response functions of the 
$(\overrightarrow{e},e^{\prime }\overrightarrow{p})$ reaction 
in the GeV region.  As such, we neglect some important physics 
such as the nuclear current conservation and the off-shell effects.  
The Dirac eikonal formalism that we used seems to agree well with 
the partial-wave expansion method in the GeV region.

Our findings are summarized as follows:

(1) The effect of the final-state interaction in $R_{TT}$ is 
caused mostly by the spin-orbit interaction, while that in  $R_T^n$ is
caused {\it solely} by the spin-orbit force.  The effect in $R_{TT}^n$ 
is caused mostly by the central force.  These are the cases for both 
of the $1p_{1/2}$-shell and the $1p_{3/2}$-shell knock-out processes 
also at both of $T_{p'}=0.515$ GeV and 3.179 GeV. 

(2) Except for the helicity-dependent $R_{LT^{\prime }}^n$, all 
normal-component responses have different signs for the $1p_{1/2}$-
and the $1p_{3/2}$-shell knock-outs.  $P_n$ thus receives different 
contributions from the two different shell-orbits. 

(3) The response functions become smaller as $Q^2$ increases, mostly 
because of the $Q^2$ dependence of the electromagnetic form factor 
of the nucleon.

(4) The contributions of the Dirac and the Pauli currents are 
equally significant 
to the response functions, but they contribute with different signs 
to the helicity-dependent response functions, $R_{LT^{\prime }}$ and $%
R_{LT^{\prime }}^n$ .

(5) The nonvanishing value of $P_n$ that is due to the the final-state 
interactions 
is insensitive to the structure of the electromagnetic current operator. 

Among these, let us make a speculative comment on (1): 
Because they vanish or almost vanish in the absence of the the spin-orbital, 
final-state interaction, detailed measurements of $R_T^n$ and $R_{TT}$ may 
reveal an interesting, spin-dependent process of the small, color-singlet 
proton that may be produced in the high-energy $(e,e^{\prime} p)$.  So far, 
no investigation has been made on the spin-dependent process except for 
a speculative description\cite{JAIN}.  It would be an interesting issue  
that may reveal more about this strange form of the proton, especially 
because most experiments are carried out in the energy region 
where the process would be incompletely controlled by the perturbative QCD. 

\begin{center}
ACKNOWLEDGMENTS
\end{center}

We would like to acknowledge Prof. J. W. Van Orden for providing us 
the response functions calculated by the partial-wave decomposition method. 
R. S. thanks Dr. W. R. Greenberg for clarifying symmetry properties of 
the response functions.
We have been benefited from the Dirac eikonal calculation 
of the high-energy ($\overrightarrow{e},e^{\prime }p$) reaction 
by Dr. A. Allder. 
This work is supported by the U. S. Department of Energy grant
at CSUN (DE-FG03-87ER40347) and by the National Science Foundation grant at
Caltech (PHY-9412818 and PHY-9420470).

\newpage
\appendix
\section{Kinematic Factors of Structure Functions}

The kinematic factors, $v$'s, in Eqs. (15) and (16) are defined to be
$v_L=\frac{Q^4}{{\bf q}^4}$, 
$v_T=\left[ \frac{Q^2}{2{\bf q}^2}+\tan \theta /2\right] $, 
$v_{TT}=\frac{Q^2}{2{\bf q}^2}$, 
$v_{LT}=\frac{Q^2}{{\bf q}^2}\left[ \frac{Q^2}{{\bf q}^2}
+\tan ^2\theta /2\right] ^{1/2}$, 
$v_{LT^{\prime }}=\frac{Q^2}{{\bf q}^2}\tan \theta /2$, 
and $v_{TT^{\prime }}=\tan \theta /2\left[ \frac{Q^2}{{\bf q}^2}
+\tan ^2\theta /2\right] ^{1/2}$ with $Q^2=-q^2.$  

The response functions are obtained by the application of 
the projection operator ${\cal P}_{{\bf a}}= |{\bf a}\rangle \langle{\bf a}|
\frac 12(1+{\bf \sigma \cdot } \widehat{{\bf a}})$  
for $\widehat{{\bf a}} = \widehat{{\bf n}},\widehat{{\bf l}}$ or 
$\widehat{{\bf t}}$).  More explicitly, they are given by 

\begin{equation}
\begin{array}{ll}
R_L=Tr\{\widetilde{R_L}{\bf I}\}, 
& R_L^n=Tr\{\widetilde{R_L}{\bf \sigma }\cdot {\bf n}\}, \\ 
R_T=Tr\{\widetilde{R_T}{\bf I}\}, 
& R_T^n=Tr\{\widetilde{R_T}{\bf \sigma }\cdot {\bf n}\}, \\ 
R_{TT}=Tr\{\widetilde{R_{TT}}{\bf I}\}/\cos 2\beta, 
& R_{TT}^n=Tr\{\widetilde{R_{TT}}{\bf \sigma }\cdot {\bf n}\}/\cos 2\beta, \\ 
R_{LT}=Tr\{\widetilde{R_{LT}}{\bf I}\}/\sin \beta, 
& R_{LT}^n=Tr\{\widetilde{R_{LT}}{\bf \sigma }\cdot {\bf n}\}/\sin \beta, \\ 
R_{LT^{\prime}}=Tr\{\widetilde{R_{LT^{\prime }}}{\bf I}\}/\cos \beta, 
& R_{LT^{\prime}}^n=Tr\{\widetilde{R_{LT^{\prime }}}
{\bf \sigma }\cdot {\bf n}\}/\cos \beta, \\ 
R_{LT}^t=Tr\{\widetilde{R_{LT}}{\bf \sigma }\cdot {\bf t}\}/\cos \beta, 
& R_{LT}^l=Tr\{\widetilde{R_{LT}}{\bf \sigma }\cdot {\bf l}\}/\cos \beta, \\ 
R_{TT}^t=Tr\{\widetilde{R_{TT}}{\bf \sigma }\cdot {\bf t}\}/\sin 2\beta, 
& R_{TT}^l=Tr\{\widetilde{R_{TT}}{\bf \sigma }\cdot {\bf l}\}/\sin 2\beta, \\ 
R_{LT^{\prime}}^t=Tr\{\widetilde{R_L}{\bf \sigma }\cdot {\bf t}\}/\sin \beta, 
& R_{LT^{\prime }}^l=Tr\{\widetilde{R_L}{\bf \sigma }\cdot {\bf l}\}
/\sin \beta, \\ 
R_{TT^{\prime}}^t=Tr\{\widetilde{R_{TT^{\prime }}}{\bf \sigma }\cdot {\bf t}\}, 
& R_{TT^{\prime }}^l=Tr\{\widetilde{R_{TT^{\prime }}}{\bf \sigma }\cdot {\bf l},
\} 
\end{array}
\end{equation}

\noindent where $\widetilde{R}$'s are given in terms of the nuclear tensor 
in the Dirac plane-wave, spinor space as 

\begin{equation}
\begin{array}{l}
\widetilde{R_L}=\overline{\omega }^{00}, \\ 
\widetilde{R_T}=\overline{\omega }^{22}+\overline{\omega }^{11}, \\ 
\widetilde{R_{TT}}=\overline{\omega }^{22}-\overline{\omega }^{11}, \\ 
\widetilde{R_{LT}}=\overline{\omega }^{20}-\overline{\omega }^{02}), \\ 
\widetilde{R_{LT^{\prime}}}=i(\overline{\omega }^{10}-\overline{\omega }^{01}), \\ 
\widetilde{R_{TT^{\prime }}}=i(\overline{\omega }^{12}-\overline{\omega }^{21}). 
\end{array}
\end{equation}

\newpage

\begin{figure}[tbp]
\caption{The coordinate system and kinematical variables of the
($\protect\overrightarrow{e} , e^{\prime} 
\protect\overrightarrow{p}$) reaction. 
The notations are the same as those used in Refs. 15 and 16.}
\label{fig1}
\end{figure}

\begin{figure}[tbp]
\caption{Response functions for the proton knocked out of the 
$1p_{1/2}$-shell of $^{16}O$ with the kinetic energy of $T_{p'}=0.515$ GeV. 
$|{\bf p} - {\bf q}|$ is the magnitude of the recoil momentum of the 
residual nucleus.  The functions calculated by the use of the the Dirac 
eikonal formalism are shown by solid curves, and those by the partial-wave 
decomposition method are shown by dotted curves.}
\label{fig2}
\end{figure}

\begin{figure}[tbp]
\caption{Unpolarized and normal-component, polarization
response functions for the proton knocked out of the 
$1p_{1/2}$-shell of $^{16}O$ with the kinetic energy of $T_{p'}=0.515$ GeV. 
$|{\bf p} - {\bf q}|$ is the magnitude of the recoil momentum of the 
residual nucleus.   
Solid curves are the DWIA results by use of the Dirac eikonal formalism, 
and dotted curves are the PWIA results.  The DWIA results with no spin-orbit 
potential ($V_{SO}=0$) are also shown in dashed curves.}
\label{fig3}
\end{figure}

\begin{figure}[tbp]
\caption{The same as the caption for Fig. 3, except for the ${\bf l}$- and 
${\bf t}$-component polarization response functions.}
\label{fig4}
\end{figure}

\begin{figure}[tbp]
\caption{The same as the caption of Fig.3, except that the proton is knocked 
out of the $1p_{3/2}$-shell.}
\label{fig5}
\end{figure}

\begin{figure}[tbp]
\caption{The same as the caption of Fig.4, except that the proton is knocked 
out with the kinetic energy of $T_{p'}=3.179$ GeV.} 
\label{fig6}
\end{figure}

\begin{figure}[tbp]
\caption{The same as the caption of Fig.5, except that the proton is knocked 
out with the kinetic energy of $T_{p'}=3.179$ GeV.} 
\label{fig7}
\end{figure}

\begin{figure}[tbp]
\caption{The same as the caption of Fig.4, except that the Dirac-type current
($F_1(q^2)\gamma^{\mu} \neq 0$ and $F_2(q^2)\gamma^{\mu} = 0$) is used. }
\label{fig8}
\end{figure}

\begin{figure}[tbp]
\caption{The same as the caption of Fig.4, except that the Pauli-type current
($F_1(q^2)\gamma^{\mu} = 0$ and $F_2(q^2)\gamma^{\mu} \neq 0$) is used. }
\label{fig9}
\end{figure}

\begin{figure}[tbp]
\caption{The normal-component polarization, $P_n$, for the proton knocked 
out of the $1p_{1/2}$- and the $1p_{3/2}$-shells with the kinetic energy of 
$T_{p'}=0.515$ GeV. The dotted curves are calculated only with the central 
potential, and the solid curves are with the full (central and spin-orbit) 
potential.} 
\label{fig10}
\end{figure}

\begin{figure}[tbp]
\caption{The same as the caption of Fig.11, except for the proton kinetic 
energy of $T_{p'}=3.179$ GeV. }
\label{fig11}
\end{figure}

\begin{references}
\bibitem{CTTH1}  A. H. Mueller, in Proceedings of the Seventeenth Rencontre
de Moriond, 1982, edited by J. Tran Thanh Van (Editions Fronti\`{e}res,
Gif-sur-Yvette, France 1982), p13.

\bibitem{CTTH2}  S. J. Brodsky, in Proceedings of the Thirteenth
International Symposium on Multiparticle Dynamics, edited by W. Kittel, 
{\it et al.} (World Scientific, Singapore, 1983).

\bibitem{CTTH3}  G. Farrar, {\it et al.}, Phys. Rev. Lett. {\bf 61}, 686 (1988).

\bibitem{CTTH4}  J. P. Ralston and B. Pire, Phys. Rev. Lett. {\bf 65}, 2343
(1990).

\bibitem{CTTH5}  B. K. Jennings and G. A. Miller, Phys. Rev. Lett. {\bf 69},
3619 (1992).

\bibitem{CTTH6}  L.Frankfurt and M. Strickman, {\it Progress in Particle and
Nuclear Physics}, {\bf 27}, 135 (1991).

\bibitem{CTTH7}  A. Kohama, K. Yazaki and R. Seki, Nucl. Phys. {\bf A551}, 
687 (1993).

\bibitem{CTEX1}  A. S. Carroll, {\it et al.}, Phys. Rev. Lett. {\bf 61}, 1698
(1988).

\bibitem{CTEX2}  N.C.R. Makins, {\it et al.}, Phys. Rev. Lett. {\bf 72}, 1986
(1994).

\bibitem{CTEX3}  R. McKeown and R. Milner, spokesmen, NE-18 (SLAC)
collaboration.

\bibitem{CTEX4}  R. McKeown, Nucl.Phys. {\bf A532}, 285c (1991).

\bibitem{GLAS}  Exp. 89-033 (spokesman, C. Glashausser).

\bibitem{SAHA}  Exp. 91-006 (spokesman, A. Saha).

\bibitem{WALLY1}  A. Picklesimer, J. W. Van Orden, and S. J. Wallace, Phys.
Rev. {\bf C32} 1312, (1985).

\bibitem{WALLY2}  A. Picklesimer and J. W. Van Orden, Phys. Rev.{\bf C35}, 
266 (1987).

\bibitem{WALLY3}  A. Picklesimer and W. Van Orden,Phys. Rev. {\bf C40}, 
290 (1989).

\bibitem{AMADO}  R. D. Amado, J. Piekarewicz, D. A. Sparrow, and J. A. McNeil,
Phys. Rev. {\bf C28}, 1663 (1983).

\bibitem{GREEN1}  W. R. Greenberg, Ph.D. Thesis, University of Washington,
1993.

\bibitem{GREEN2}  W. R. Greenberg and G. A. Miller, Phys. Rev. {\bf C49},
2747 (1994).

\bibitem{ALDER}  A. Alder, Ph.D. Thesis, California Institute of Technology,
1992.

\bibitem{BIAN}  A. Bianconi and M. Radici, Phys. Lett. {\bf B363}, 24 (1995). 

\bibitem{BIAN2}  A. Bianconi and M. Radici, Phys. Rev.{\bf C53}, R563 (1996). 

\bibitem{SEROT}  B. D. Serot and J. D. Walecka, Adv. Nucl. Phys. {\bf 16},
327 (1986).

\bibitem{BEN}  O. Benhar, {\it et al.}, Phys. Rev. Lett. {\bf 69}, 881 (1992).

\bibitem{SEKI}  R. Seki, T. D. Shoppa, A. Kohama, and K. Yazaki, Phys. Lett. 
{\bf B383}, 133 (1996); N. N. Nikolaev, {\it et al.}, Phys. Lett. {\bf B317}, 
281 (1993).

\bibitem{DONN}  T. W. Donnelly and J. D. Walecka, Annu. Rev. Nucl. Sci. {\bf %
25} 329 (1975).

\bibitem{BDREL}  J. D. Bjorken and S. D.Drell, {\it Relativistic Quantum
Mechanics} (McGraw-Hill, New York, 1964).

\bibitem{FORE}  T. D. Forest, Jr, Nucl. Phys. {\bf A392}, 232 (1983).

\bibitem{POLLO}  S. Pollock, H. W. L. Naus and J. H. Koch, Phys. Rev. {\bf %
C53}, 2304 (1996).

\bibitem{WALLACE}  J. A. Wallace, J. R. Shepard and S. J. Wallace, Phys.
Rev. Lett. {\bf 50}, 1439 (1983).

\bibitem{HOSHIZ}  N. Hoshizaki, Prog. Theor. Phys. {\bf 60}, 1796 (1978).

\bibitem{HIGUCH}  Y. Higuchi and N. Hoshizaki, Prog. Theor. Phys. {\bf 62},
849 (1979).

\bibitem{HOHLER}  G. H\"ohler, {\it et al.}, Nucl. Phys. {\bf B114}, 505 (1976).

\bibitem{WPRI} Private communication with J. W. Van Orden.

\bibitem{JAIN} P. Jain, B. Pire, and J. P. Ralston, Phys. Rep. {\bf 271}, 
67 (1996).


\end{references}
\end{document}